 \documentclass{ws-ijmpa}

\usepackage[super,compress]{cite}
\usepackage{graphicx}

\begin{document}
\markboth{Livia Terlizzi}{Studies on environment-friendly gas mixtures for the Resistive Plate Chambers of the ALICE Muon Identifier}

\catchline{}{}{}{}{}
%
%%%%%%%%%%%%%%%%%%%%%%%%%%%%%%%%%%%%%%%%%%%%%%%%%%%%%%%%%%%%%%%%%%%%

\title{
	Studies on environment-friendly gas mixtures for the Resistive Plate Chambers of the ALICE Muon Identifier 
}

\author{Livia Terlizzi$^k$ on behalf of the ALICE Collaboration and ECOGAS Collaboration:\\ M. Abbrescia$^g$, G. Aielli$^b$, G. Alberghi$^c$, M. C. Arena$^q$, M. Barroso$^p$, L. Benussi$^d$, A. Bianchi$^k$, S. Bianco$^d$, D. Boscherini$^c$, A. Bruni$^c$, P. Camarri$^b$, R. Cardarelli$^a$, M. Corbetta$^n$,\\ A. Di Ciaccio$^b$, L. Congedo$^g$, M. De Serio$^g$, L. Di Stante$^b$, P. Dupieux$^l$, J. Eysermans$^j$,\\ A. Ferretti$^k$ M. Ferrini$^e$, M. Gagliardi$^k$, G. Galati$^g$, A. Gelmi$^g$, R. Guida$^n$, B. Joly$^l$, B. Liberti$^a$, B. Mandelli$^n$, S.P. Manen$^l$, L. Massa$^c$, P. Mereu$^k$, L. Micheletti$^k$, L. Passamonti$^d$, A. Pastore$^r$, E. Pastori$^a$, D. Piccolo$^d$, D. Pierluigi$^d$, A. Polini$^c$, G. Proto$^a$, G. Pugliese$^g$, E. Puleo$^k$, \\ L. Quaglia$^k$, G. Rigoletti$^{m,n}$, M. Romano$^c$, A. Russo$^d$, A. Samalan$^i$, P. Salvini$^h$, R. Santonico$^b$, G. Saviano$^e$, S. Simone$^g$, M. Tytgat$^i$, E. Vercellin$^k$, M. Verzeroli$^q$ and N. Zaganidis$^o$ \\}

\address{$^a$INFN, Tor Vergata, Rome, Italy \\
		$^b$Dipartimento di Fisica di Roma Tor Vergata, Rome, Italy \\
		$^c$INFN, Bologna, Italy \\
		$^d$Laboratori Nazionali di Frascati dell’INFN, Italy \\
		$^e$Sapienza Universita di Roma, Dipartimento di Ingegneria Chimica Materiali Ambiente, Rome, Italy \\
		$^f$Laboratori Nazionali di Frascati dell’INFN, Italy \\
		$^g$Dipartimento di Fisica di Bari and Sezione INFN di Bari, Italy \\
		$^h$Sezione INFN di Pavia, Italy \\
		$^i$Ghent University, Dept. of Physics and Astronomy, Proeftuinstraat 86, B-9000 Ghent, Belgium \\
		$^j$MIT, Cambridge, Massachusetts, USA \\
		$^k$Università di Torino and Sezione INFN di Torino, Via Giuria 1, 10125 Torino, Italy \\
		$^l$Clermont Université, Université Blaise Pascal, CNRS/IN2P3, Laboratoire de Physique \\
		Corpusculaire, BP 10448, F-63000 Clermont-Ferrand, France \\
		$^m$Université Claude Bernard Lyon I, Lyon, France \\
		$^n$CERN, Geneve, Switzerland \\
		$^o$Univ. Iberoamericana, Mexico City, Mexico \\
		$^p$Universidade do Estado do Rio de Janeiro \\
		$^q$Università degli Studi di Pavia, Italy \\
		$^r$Sezione INFN di Bari, Italy \\		
		E-mail: livia.terlizzi@cern.ch}

\maketitle

\begin{history}
\received{Day Month Year}
\revised{Day Month Year}
\end{history}

\begin{abstract}
ABSTRACT: Due to their simplicity and comparatively low cost, Resistive Plate Chambers are gaseous detectors widely used in high-energy and cosmic rays physics, when large detection areas are needed. However, the best gaseous mixtures are currently based on tetrafluoroethane, which has the undesirable characteristic of a large Global Warming Potential (GWP) of about 1400 and, because of this, it is currently being phased out from industrial use. Tetrafluoropropene (which has a GWP close to 1) is being considered as a possible replacement.
Since tetrafluoropropene is more electronegative than tetrafluoroethane, it has to be diluted with gases with a lower attachment coefficient in order to maintain the operating voltage close to 10 kV. One of the main candidates for this role is carbon dioxide. In order to ascertain the feasibility and the performance of tetrafluoropropene-CO2 based mixtures, an R$\&$D program is being carried out within the ALICE collaboration, employing an array of 72 Bakelite RPCs (Muon IDentifier, MID) in order to identify muons. Different proportions of tetrafluoropropene and CO2, with the addition of small quantities of isobutane and sulphur hexafluoride, have been tested with 50x50 cm$^2$ RPC prototypes with 2 mm wide gas gap and 2 mm thick Bakelite electrodes.
In this contribution, results from tests with cosmic rays will be presented, together with data concerning the current drawn by a RPC exposed to the gamma-ray flux of the Gamma Irradiation Facility (GIF) at CERN.

\end{abstract}

%\ccode{PACS numbers:}

%\tableofcontents

\section{The ALICE Muon Identification System (MID)}

\begin{figure}[tb]
	\centerline{\includegraphics[width=7cm]{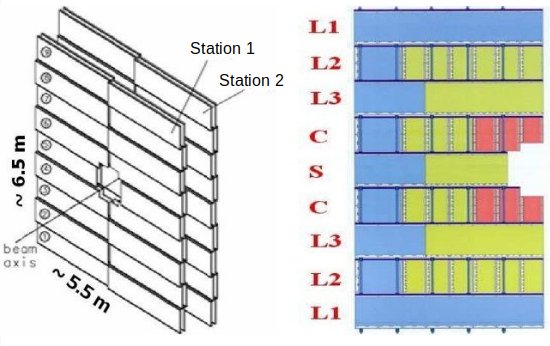}}
	\caption{A schematic view of the ALICE MID (left) and detector composition of an half plane (right). {\it L, C} and {\it S} refer to the differents geometries of the RPCs: {\it L1, L2} and {\it L3} indicate the {\it Long} RPCs, {\it C} the {\it Cut} ones and {\it S} the {\it Short} one. The colours indicate the different strip segmentation. \label{midlayout1}}
\end{figure}

The ALICE (A Large Ion Collider Experiment)\cite{ALICE1} at CERN is one of the main experiments at the Large Hadron Collider (LHC) and it is designed to study proton-proton and heavy-ion collisions (such as Pb-Pb) at ultra-relativistic energies. \\ 
The main goal of ALICE is to assess the properties of the Quark Gluon Plasma (QGP), which is a state of matter where quarks and gluons are de-confined. Among the main observables used to study the QGP there is the production of heavy quarkonia, i.e. $c \overline{c}$ and $b \overline{b}$ bound states. The presence of QGP modifies the quarkonium production rates in heavy-ion collisions via the competing processes of suppression by colour screening and regeneration by quark recombination\cite{QGP1} . \\
For this reason ALICE is equipped with a forward Muon Spectrometer\cite{MS} which detects quarkonia via their di-muon decay channel in the rapidity interval $2.5 < \eta < 4$. It is composed of a hadron absorber, a dipole magnet, a five-station tracking system and a Muon IDentifier (MID) placed downstream a 120 cm thick iron wall. \\ 
The MID system is composed of 72 Resistive Plate Chambers (RPCs)\cite{MID1} , which are position-sensitive gaseous detectors with a spatial resolution of the order of few mm and time resolution of the order of ns. The MID RPCs are arranged in two stations of two planes each. The two stations are situated at 16.1 m and 17.1 m from the Interaction Point (IP). Each station consists of 18 RPC modules and covers an area of 5.5 $\times$ 6.5 m$^2$ with a 1.2 $\times$ 1.2 m$^2$ central hole to accomodate the beam pipe and its shielding. A single RPC size is 70 $\cdot$ 270 cm$^2$  and there are three different geometry shapes: Long ({\it L}), Cut ({\it C}) and Short ({\it S}). Each RPC is equipped with orthogonal copper X-Y read-out strips in order to collect spatial information along both {\it x} and {\it y} directions, on the plane perpendicular to the beam axis. There are $\sim$21k strips, with pitches (1, 2 and 4 cm wide) and lengths increasing with their distance from the beam axis, in order to provide an almost flat occupancy throughout all the plane surface and to keep the momentum resolution roughly constant (see Fig.~\ref{midlayout1}).

\subsection{ALICE MID Resistive Plate Chambers}

The ALICE MID RPCs\cite{ALICERPC} are single-gap (2 mm thick) detectors with resistive bakelite electrodes (2 mm thick, $\rho$ $\simeq$ 3 $\cdot$ 10$^9$ - 1 $\cdot$ 10$^{10}$ $\Omega$ cm). The detector installation in the ALICE cavern took place in 2007 and the RPCs have been operating since the beginning of the Run 1 in 2010. During Run 1 and Run 2, the signal was discriminated by a front-end electronics\cite{Frontend} with a threshold value of 7 mV and without any pre-amplification stage. \\ The ALICE RPCs worked with an effective applied HV of about 10.2 - 10.5 kV at 970 mbar of pressure and 20$^\circ$C. They worked in the so called maxi-avalanche mode with an average charge per hit of 100 pC, while the maximum rate capability was about 100 Hz$/$cm$^2$ \cite{Ageingtest1}. \\ 
From Run 3 onwards, to cope with the increased collision rate, the ALICE MID RPCs will work in avalanche mode in order to reduce the charge and increase the rate capability. This will be possible thanks to new front-end electronics (FEERIC ASIC\cite{FEERIC1}) which includes a pre-amplification stage of the analogue signal before discrimination, with a Q$_{\mathrm{induced}}$ threshold of 130 fC. The effective HV applied will be between 9.7 kV and 10 kV. \\
Regarding the gas mixture, during Run 3 the RPCs will be flushed with the same gas mixture as that used during Run 1 and Run 2: 89.7$\%$ tetrafluoroethane (C$_{2}$H$_{2}$F$_{4}$, R134a) 10$\%$ isobutane ({\it i}-C$_{4}$H$_{10}$), and 0.3$\%$ sulphur hexafluoride (SF$_{6}$). The mixture is humidified at 35-40$\%$ RH in order to avoid any variations in the resistivity of the bakelite electrodes. \\
This gas mixture is flammable because of the presence of isobutane, while tetrafluoroethane and sulphur hexafluoride are greenhouse gases, making the current gas mixture used for the RPCs not environmentally friendly. Although the total gas volume of the MID is small ($\sim$ 0.3 m$^3$), a more environmentally friendly gas mixture would be welcomed, so R$\&$D studies on mixtures with lower GWP have been performed (and are still ongoing) and the results obtained up to now are shown in this paper.

\begin{figure}[tb]
	\centerline{\includegraphics[width=7.5cm]{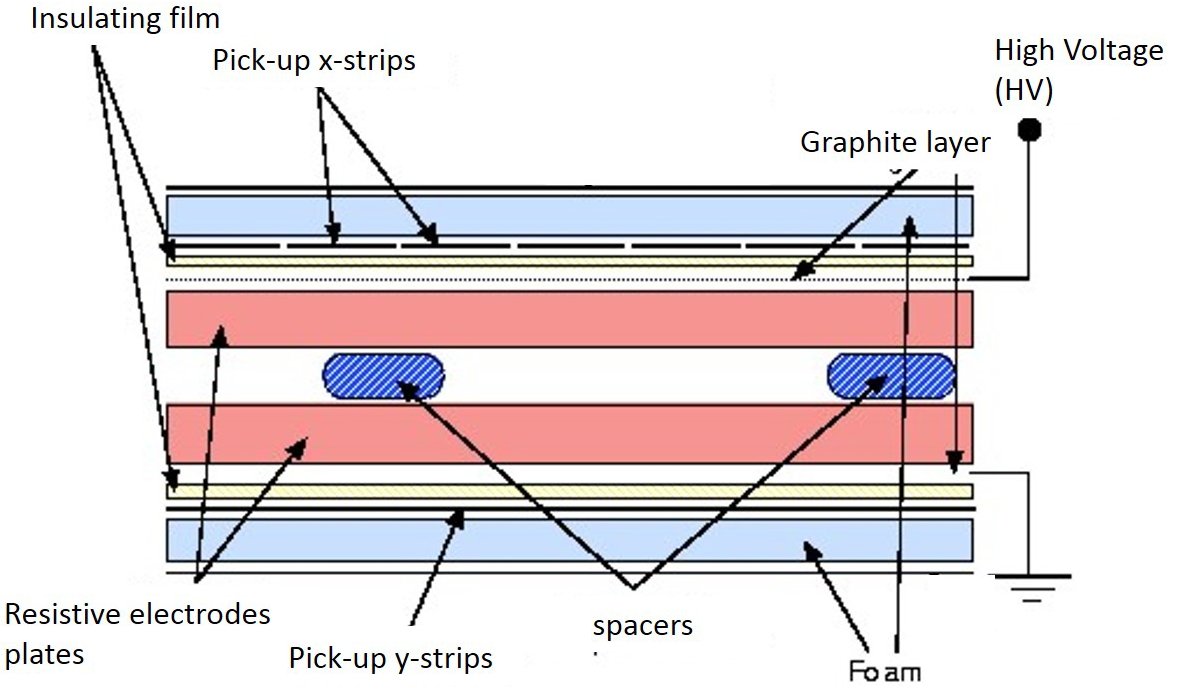}}
	\caption{A schematic view of an ALICE RPC detector \label{RPCschematic}}
\end{figure}

\section{Search for environment-friendly gas mixture}

\subsection{Reasons for R$\&$D studies}

European regulations\cite{European} have imposed a reduction of the emission into the atmosphere of fluorinated Greenhouse Gases (GHG), e.g. tetrafluoroethane. This leads to a driving up of the procurement costs. Therefore, CERN is pushing for a search for more environmentally friendly gas mixtures for particle detectors, in order to limit the GHG emissions in atmosphere, and also for economic reasons. \\ Tetrafluoroethane and sulphur hexafluoride are greenhouse gases and they are classified according to their Global Warming Potential (GWP), which is the measure of how much energy the emissions of 1 ton of a gas will absorb over a given period of time, relative to the emissions of 1 ton of carbon dioxide (CO$_{2}$). The GWP$_{100yr}$ of tetrafluoroethane is 1300, while that of SF$_6$ is 23500\cite{IPCC} . Considering that the GWP$_{100yr}$ of isobutane is 3, and taking into account the percentage of each gas in the mixture, the GWP of the current ALICE gas mixture is equal to 1351. \\ 
The ALICE MID group started to study new environmentally friendly and non-flammable gas mixtures in 2017, with a cosmic rays set-up (see subsection \ref{cosmic}). In 2018 a collaboration, called ECOGAS, between the RPC groups of ALICE, CMS and ATLAS, together with the CERN gas team, was started. In 2019 the SHiP collaboration joined the team and R$\&$D studies are still ongoing. The goal of these R$\&$D studies is to determine a new low GWP gas mixture that provides similar detector performances in terms of efficiency and streamer probability with respect to the current RPCs gas mixture. \\

\subsection{From tetrafluoroethane to tetrafluoropropene}

Up to now work is being carried out on a replacement for tetrafluoroethane, since R134a is responsible for 95$\%$ of the total GWP of the RPC gas mixtures at the LHC. \\ An appropriate candidate may be found among Hydro-Fluoro-Olefin (HFO) gases, since they have similar properties to the R134a but with a low GWP. The tetrafluoropropene C$_3$H$_2$F$_4$ (trade-name HFO-1234ze) was chosen \cite{Guida} \cite{Liberti} \cite{Bianchi1} . This gas is not flammable at room temperature and it has a GWP lower than 1. \\
However, a direct replacement of tetrafluoroethane with tetrafluoropropene is not possible, since electron capture is more dominant for C$_3$H$_2$F$_4$ than for C$_2$H$_2$F$_4$. This would lead to an operating voltages for RPCs higher than 15 kV, which is not advisable for our system. \\
Several studies \cite{Bianchi2} have been carried out with the addition of different gases to the tetrafluoropropene-based gas mixture, in order to lower the HV Working Point (WP). The most promising so far is the gas mixture with the addition of CO$_2$. \\ In this paper results will be shown on HFO-based gas mixtures with different percentages of CO$_2$, {\it i}-C$_4$H$_{10}$ and SF$_6$. 

\subsection{Cosmic rays experimental set-up \label{cosmic}}

\begin{figure}[tb]
	\centerline{\includegraphics[width=7cm]{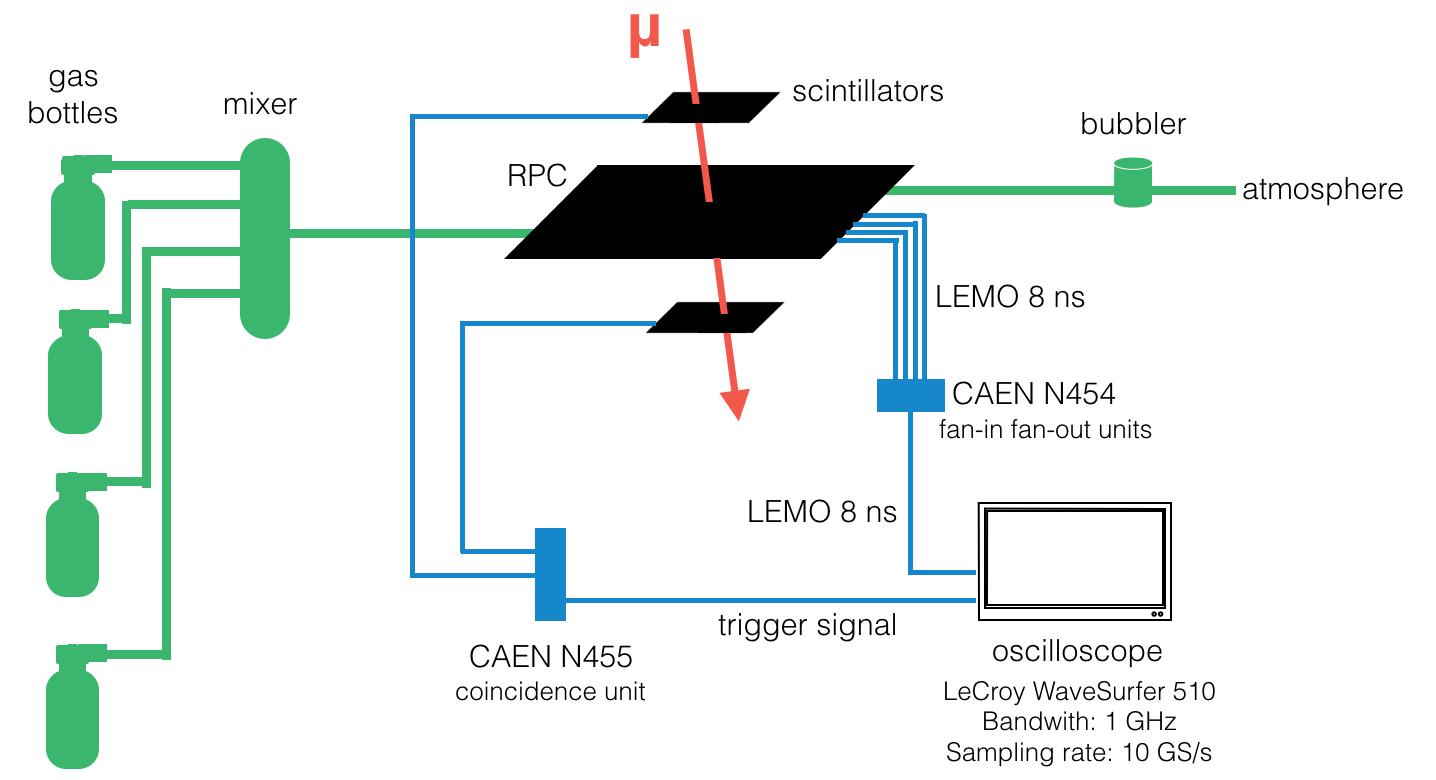}}
	\caption{Schematic view of the experimental set-up for R$\&$D studies on eco-friendly gas mixtures at the INFN Torino laboratory. \label{exp_setup}}
\end{figure}

The tests on eco-friendly gas mixtures are performed in the INFN Torino laboratory, using a cosmic ray test station (see figure \ref{exp_setup}). \\ These studies are perfomed with a small size ALICE RPC (50 $\times$ 50 cm$^2$ size, 2mm gas gap, 16 readout strips per side with 2 cm pitch) placed horizontally and exposed to cosmic rays. For the trigger three plastic scintillators with a total area of about 6 $\times$ 6 cm$^2$) are used. \\ The detector is equipped on one strip end with the ALICE FEERIC front-end electronics, which amplifies and discriminates signals (threshold for the data acquisition is Q$_\mathrm{{induced}}$ = 130 fC, 70 mV after amplification, the same as the one used for FEERIC installed in ALICE \cite{FEERICALICE}). On the other strip end, the analogue signals are summed with a fan-in/fan-out module and digitized by an oscilloscope. In this way it is possible to measure the signal amplitude and charge. \\
The HV is applied with temperature and pressure correction\footnote{All the results in this paper are given with the corrected high voltage HV: \begin{equation}
	HV = HV_\mathrm{{app}} \frac{p_0}{p} \frac{T}{T_0}
	\label{HV_correct}
	\end{equation} where p$_0$ and T$_0$ are reference values equal to 1000 mbar and 293.15$^\circ$K respectively, while p and T are the pressure and the temperature in the laboratory.}.
Moreover, there is the possibility to mix up to 4 different gases, using a dedicated gas mixing unit, and the humidity of the gas mixture is kept constant at 35-40$\%$ RH in order to avoid variations of the electrode resistivity. \\

\section{Tetrafluoropropene - based gas mixture}

\begin{figure}[b]
	\centerline{\includegraphics[width=7cm]{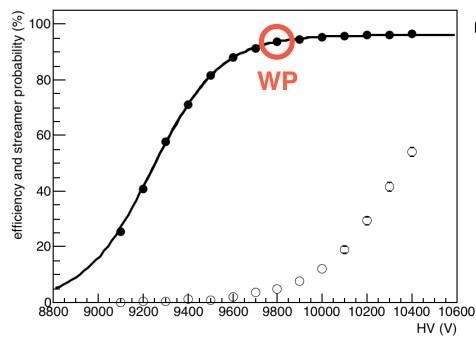}}
	\caption{Efficiency curve (black dots) and streamer probability (white dots) of a RPC detector flushed with standard ALICE gas mixture. The solid line is a sigmoid-function fit to the efficiency curve. \label{standard}}
\end{figure}

The ALICE standard gas mixture is used as a reference, so, first of all, the efficiency curve versus HV and the streamer probability have been evaluated by flushing the RPC with the standard ALICE mixture (89.7$\%$ C$_2$H$_2$F$_4$, 10 $\%$ {\it i}-C$_4$H$_{10}$ and 0.3$\%$ SF$_6$). \\
The plot in figure \ref{standard} shows the efficiency and the streamer probability for this mixture. As said before, the new eco-friendly gas mixture should provide similar detector performance with respect to the standard one, without impacting negatively on the detector working parameters, especially on the HV working point WP, which corresponds to the beginning of the efficiency plateau and in this case is 9.8kV. The streamer probability at the working point is 5$\%$. The occurrence of streamers in the ALICE RPC has to be kept to a minimum because it leads both to a reduction in the rate capability and an increase in the cluster size, and may enhance ageing effects\cite{ageingALICE} . 

\begin{figure}[tb]
	\centerline{\includegraphics[width=6cm]{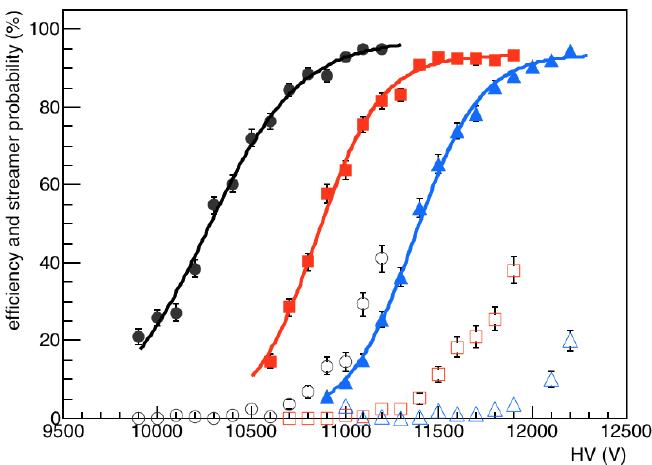}}
	\caption{Efficiency curves and streamer probability for gas mixtures with different ratios between C$_3$H$_2$F$_4$ and CO$_2$ ({\it i}-C$_4$H$_{10}$ and SF$_6$ are kept constant at 10$\%$ and 1$\%$ respectively): (i) 33.5$\%$ C$_3$H$_2$F$_4$ and 55.5$\%$ CO$_2$ (black dots), (ii) 39$\%$ C$_3$H$_2$F$_4$ and 50$\%$ CO$_2$ (red squares), (iii) 44.5$\%$ C$_3$H$_2$F$_4$ and 44.5$\%$ CO$_2$ (blue triangles). \label{CO2}}
\end{figure}

The gas mixtures tested in this paper are all made up from four different gases and, since the contribution of each gas in the composition is quite complex to investigate, we change the proportions of only two gases at a time. \\ \newpage First, efficiency and streamer probability have been measured for tetrafluoropropene based gas mixtures, with the addition of different concentrations of CO$_2$. In figure \ref{CO2} mixtures with different C$_3$H$_2$F$_4$ and CO$_2$ ratio are shown, while {\it i}-C$_4$H$_{10}$ and SF$_6$ are kept constant. By increasing C$_3$H$_2$F$_4$ and decreasing CO$_2$ the WP is shifted towards higher voltages, while there is no significant variation of the streamer probability at the WP.

\begin{figure}[b]
	\centerline{\includegraphics[width=6cm]{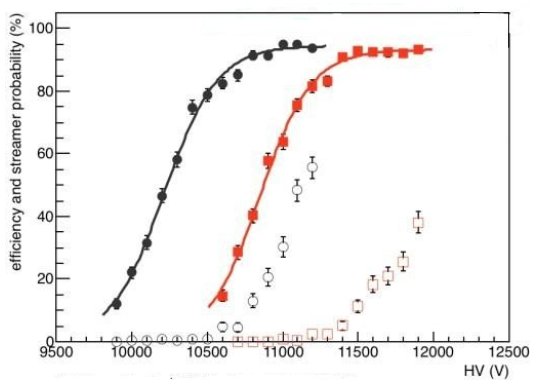}}
	\caption{Efficiency curves and streamer probability for gas mixtures with different ratios between C$_3$H$_2$F$_4$ and {\it i}-C$_4$H$_{10}$ (CO$_2$ and SF$_6$ are kept constant at 50$\%$ and 1$\%$ respectively): (i) 29$\%$ C$_3$H$_2$F$_4$ and 20$\%$ {\it i}-C$_4$H$_{10}$ (black dots), (ii) 39$\%$ C$_3$H$_2$F$_4$ and 10$\%$ {\it i}-C$_4$H$_{10}$ (red squares). \label{iso}}
\end{figure}

In order to investigate the effects of the isobutane, we have also measured efficiency and streamer probability for tetrafluoropropene-based gas mixtures,
with the addition of different concentrations of isobutane. As shown in figure \ref{iso}, we can see a similar behaviour as for CO$_2$, leading to the conclusion that there is a strong dependence between the concentration of C$_3$H$_2$F$_4$ and the WP, due to the higher electron affinity of tetrafluoropropene with respect to tetrafluoroethane. 

\begin{figure}[b]
	\centerline{\includegraphics[width=6cm]{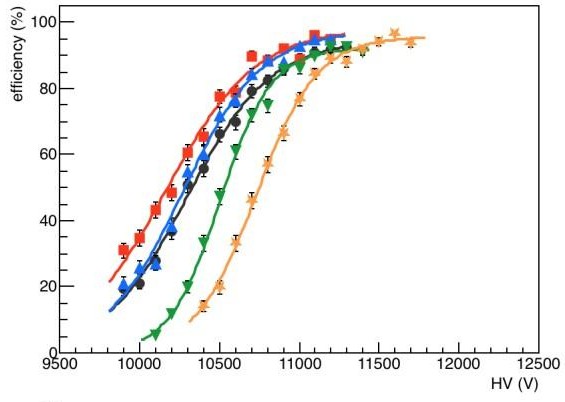}}
	\caption{Efficiency curves for gas mixtures with different ratios between CO$_2$ and {\it i}-C$_4$H$_{10}$ (C$_3$H$_2$F$_4$ and SF$_6$ are kept constant at 34$\%$ and 1$\%$ respectively): (i) 65.5$\%$ CO$_2$ and 0$\%$ {\it i}-C$_4$H$_{10}$ (black dots), (ii) 60.5$\%$ CO$_2$ and 5$\%$ {\it i}-C$_4$H$_{10}$ (red squares), (iii) 55.5$\%$ CO$_2$ and 10$\%$ {\it i}-C$_4$H$_{10}$ (blue triangles), (iv) 50$\%$ CO$_2$ and 15$\%$ {\it i}-C$_4$H$_{10}$ (green reverse triangles), (v) 44.5$\%$ CO$_2$ and 20$\%$ {\it i}-C$_4$H$_{10}$ (orange stars). \label{isoCO2}}
\end{figure}

\begin{figure}[b]
	\centerline{\includegraphics[width=6cm]{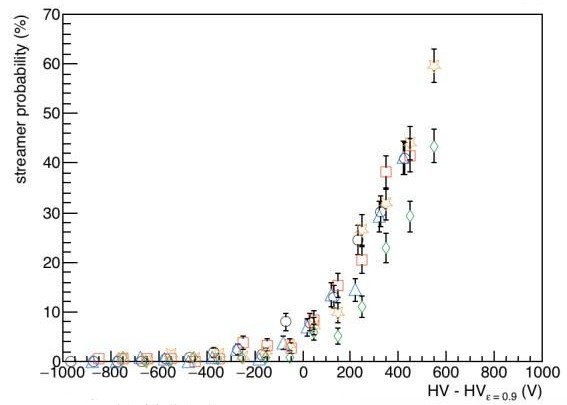}}
	\caption{Streamer probability for gas mixtures with different ratios between CO$_2$ and {\it i}-C$_4$H$_{10}$ (C$_3$H$_2$F$_4$ and SF$_6$ are kept constant at 34$\%$ and 1$\%$ respectively): (i) 65.5$\%$ CO$_2$ and 0$\%$ {\it i}-C$_4$H$_{10}$ (black dots), (ii) 60.5$\%$ CO$_2$ and 5$\%$ {\it i}-C$_4$H$_{10}$ (red squares), (iii) 55.5$\%$ CO$_2$ and 10$\%$ {\it i}-C$_4$H$_{10}$ (blue triangles), (iv) 50$\%$ CO$_2$ and 15$\%$ {\it i}-C$_4$H$_{10}$ (green rhombuses), (v) 44.5$\%$ CO$_2$ and 20$\%$ {\it i}-C$_4$H$_{10}$ (orange stars). \label{streamer_isoCO2}}
\end{figure}

\newpage
Since isobutane affects the gas mixture flammability, its reduction or, preferably, removal is desirable for safety reasons. In figures \ref{isoCO2} and \ref{streamer_isoCO2} the efficiency curves and the streamer probability for gas mixtures with different CO$_2$ and {\it i}-C$_4$H$_{10}$ ratio, with constant C$_3$H$_2$F$_4$ (34$\%$) and SF$_6$ (1$\%$), are shown. The WP does not vary monotonically with the ratio, decreasing by 100 V when raising the isobutane fraction from 0$\%$ to 5$\%$ and increasing by 500 V when the isobutane fraction is 20$\%$, while there are very similar streamer probability in all cases (note that Fig. \ref{streamer_isoCO2} refers to a shift between the applied HV and the HV at the 90$\%$ of efficiency). Although the reduction of isobutane is desirable, its reduction also results in a less steep turn-on of the efficiency curve. \\

\newpage

\begin{figure}[b]
	\centerline{\includegraphics[width=6.5cm]{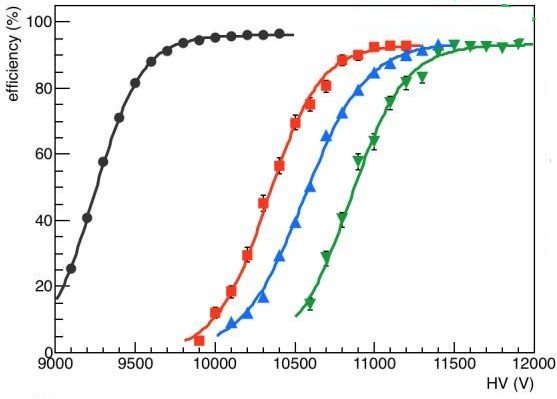}}
	\caption{Efficiency curves for gas mixtures with different percentage of SF$_6$, compared with ALICE standard gas mixture (i) 0.3$\%$ SF$_6$, 89.7$\%$ C$_2$H$_2$F$_4$, 10$\%$ {\it i}-C$_4$H$_{10}$ (black dots). (ii) 0.3$\%$ SF$_6$, 39.7$\%$ C$_3$H$_2$F$_4$, 50$\%$ CO$_2$, 10$\%$ {\it i}-C$_4$H$_{10}$ (red squares), (iii) 0.6$\%$ SF$_6$, 39.4$\%$ C$_3$H$_2$F$_4$, 50$\%$ CO$_2$, 10$\%$ {\it i}-C$_4$H$_{10}$ (blue triangles), (iv) 1.0$\%$ SF$_6$, 39.3$\%$ C$_3$H$_2$F$_4$, 50$\%$ CO$_2$, 10$\%$ {\it i}-C$_4$H$_{10}$ (green reverse triangles).\label{SF6}}
\end{figure}

\begin{figure}[b]
	\centerline{\includegraphics[width=6.5cm]{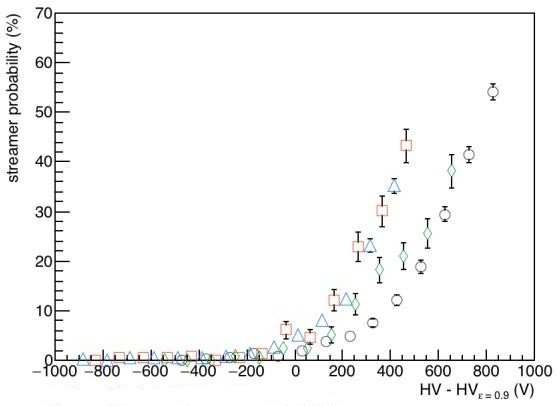}}
	\caption{Streamer probability for gas mixtures with different percentage of SF$_6$, compared with ALICE standard gas mixture (i) 0.3$\%$ SF$_6$, 89.7$\%$ C$_2$H$_2$F$_4$, 10$\%$ {\it i}-C$_4$H$_{10}$ (black dots). (ii) 0.3$\%$ SF$_6$, 39.7$\%$ C$_3$H$_2$F$_4$, 50$\%$ CO$_2$, 10$\%$ {\it i}-C$_4$H$_{10}$ (red squares), (iii) 0.6$\%$ SF$_6$, 39.4$\%$ C$_3$H$_2$F$_4$, 50$\%$ CO$_2$, 10$\%$ {\it i}-C$_4$H$_{10}$ (blue triangles), (iv) 1.0$\%$ SF$_6$, 39.3$\%$ C$_3$H$_2$F$_4$, 50$\%$ CO$_2$, 10$\%$ {\it i}-C$_4$H$_{10}$ (green rhombuses). \label{streamerSF6}}
\end{figure}

The role of the SF$_6$ in the gas mixture is to suppress the streamers\cite{SF6}. Given this, we have studied tetrafluoropropane-based gas mixtures with different percentage of SF$_6$ and we have compared them with the ALICE standard gas mixture. In figures \ref{SF6} and \ref{streamerSF6} the efficiency curves and the streamer probability for these gas mixtures are shown. A small variation of SF$_6$, from 0.3$\%$ to 1.0$\%$, leads to a variation of the WP of $\sim$ 500 V, while there is no significant variation in the streamer probability when increasing SF$_6$ from 0.3$\%$ to 0.6$\%$. The suppression of the streamers is slightly higher with 1.0$\%$ of SF$_6$. \\

\newpage
\subsection{Most promising gas mixtures (up to now)}

\begin{figure}[b]
	\centerline{\includegraphics[width=6.5cm]{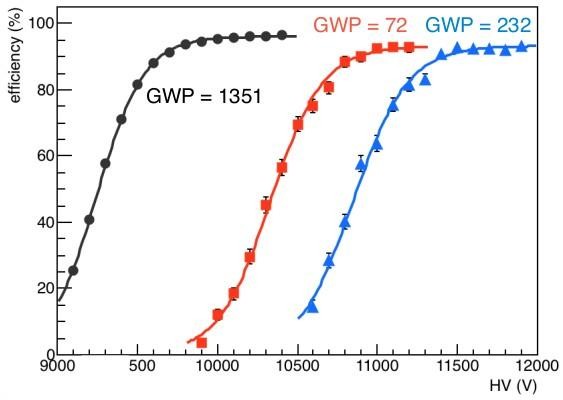}}
	\caption{Efficiency curves for the two most promising gas mixtures (up to now) compared with the ALICE standard gas mixture (i) (black dots). (ii) 39.7$\%$ C$_3$H$_2$F$_4$, 50$\%$ CO$_2$, 10$\%$ {\it i}-C$_4$H$_{10}$, 0.3$\%$ SF$_6$ (red squares), (iii) 39.0$\%$ C$_3$H$_2$F$_4$, 50$\%$ CO$_2$, 10$\%$ {\it i}-C$_4$H$_{10}$, 1.0$\%$ SF$_6$ (blue triangles).\label{promising}}
\end{figure}

\begin{figure}[b]
	\centerline{\includegraphics[width=6.5cm]{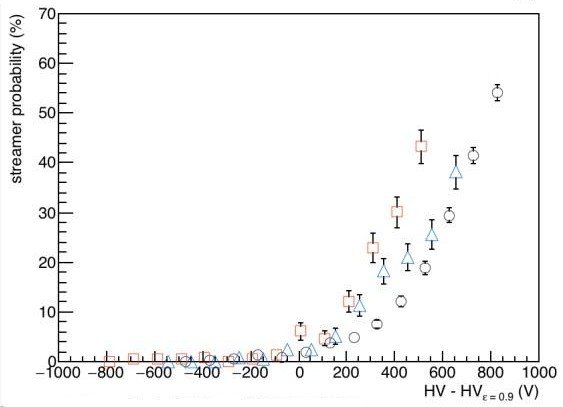}}
	\caption{Streamer probability for the two most promising gas mixtures (up to now) compared with the ALICE standard gas mixture (i) (black dots). (ii) 39.7$\%$ C$_3$H$_2$F$_4$, 50$\%$ CO$_2$, 10$\%$ {\it i}-C$_4$H$_{10}$, 0.3$\%$ SF$_6$ (red squares), (iii) 39.0$\%$ C$_3$H$_2$F$_4$, 50$\%$ CO$_2$, 10$\%$ {\it i}-C$_4$H$_{10}$, 1.0$\%$ SF$_6$ (blue triangles). \label{streamerpromising}}
\end{figure}

Given the results shown above, we concluded by defining two promising gas mixtures. In figures \ref{promising} and \ref{streamerpromising} the efficiency curves and the streamers probability for these two gas mixtures, compared with the ALICE standard mixture, are shown. \\
The one with 39.7$\%$ C$_3$H$_2$F$_4$, 50$\%$ CO$_2$, 10$\%$ {\it i}-C$_4$H$_{10}$, 0.3$\%$ SF$_6$ has a GWP = 72 ( $\sim$ 20 times lower than the
ALICE mixture), but the WP is $\sim$ 1 kV higher. The streamer probability is also higher than that with the ALICE mixture. \\
The one with 39.0$\%$ C$_3$H$_2$F$_4$, 50$\%$ CO$_2$, 10$\%$ {\it i}-C$_4$H$_{10}$, 1.0$\%$ SF$_6$ has a GWP = 232 ( $\sim$ 5 times lower than the
ALICE mixture) and a WP $\sim$ 1.5 kV higher than the standard mixture. The streamer probability is similar to that with the ALICE mixture, because of the higher percentage of SF$_6$. 

\newpage
\section{R$\&$D at Gamma Irradiation Facility (GIF) at CERN}

\subsection{Gamma Irradiation Facility (GIF++)}

Finding a new eco-friendly gas mixtures with suitable performance is not enough. It is also important to perform an ageing test on the new gas mixtures in order to see if they allow for stable long-term operations. Ageing is strongly influenced by the chemical composition of the gas mixture: in the ALICE MID RPCs ageing is mainly due to a deterioration of the electrodes surface smoothness probably due to UV and chemical action. This causes the formation of tips and leads to an increase of the dark current\cite{Quaglia} . \\ For this reason we tested RPCs under gamma irradiation to simulate many years of operations at the LHC while keeping track of current drawn over time. These tests are performed at the CERN Gamma Irradiation Facility (GIF++). ECOgas@GIF++ collaboration (ALICE, ATLAS, CMS, EP-DT, SHiP) has been created to perform these aging tests on HFO-based gas mixture with low GWP. RPCs have been provided by the different groups and have been installed onto a common trolley placed inside the bunker at GIF++. The GIF is equipped with a Cs137 radioactive source (14 TBq), emitting 662 keV gamma rays. Filters are used to modulate the radiation on the detectors under test and there is the possibility to have muon beam for beam tests. \\ 

\subsection{Ageing test}

\begin{figure}[tb]
	\centerline{\includegraphics[width=8.5cm]{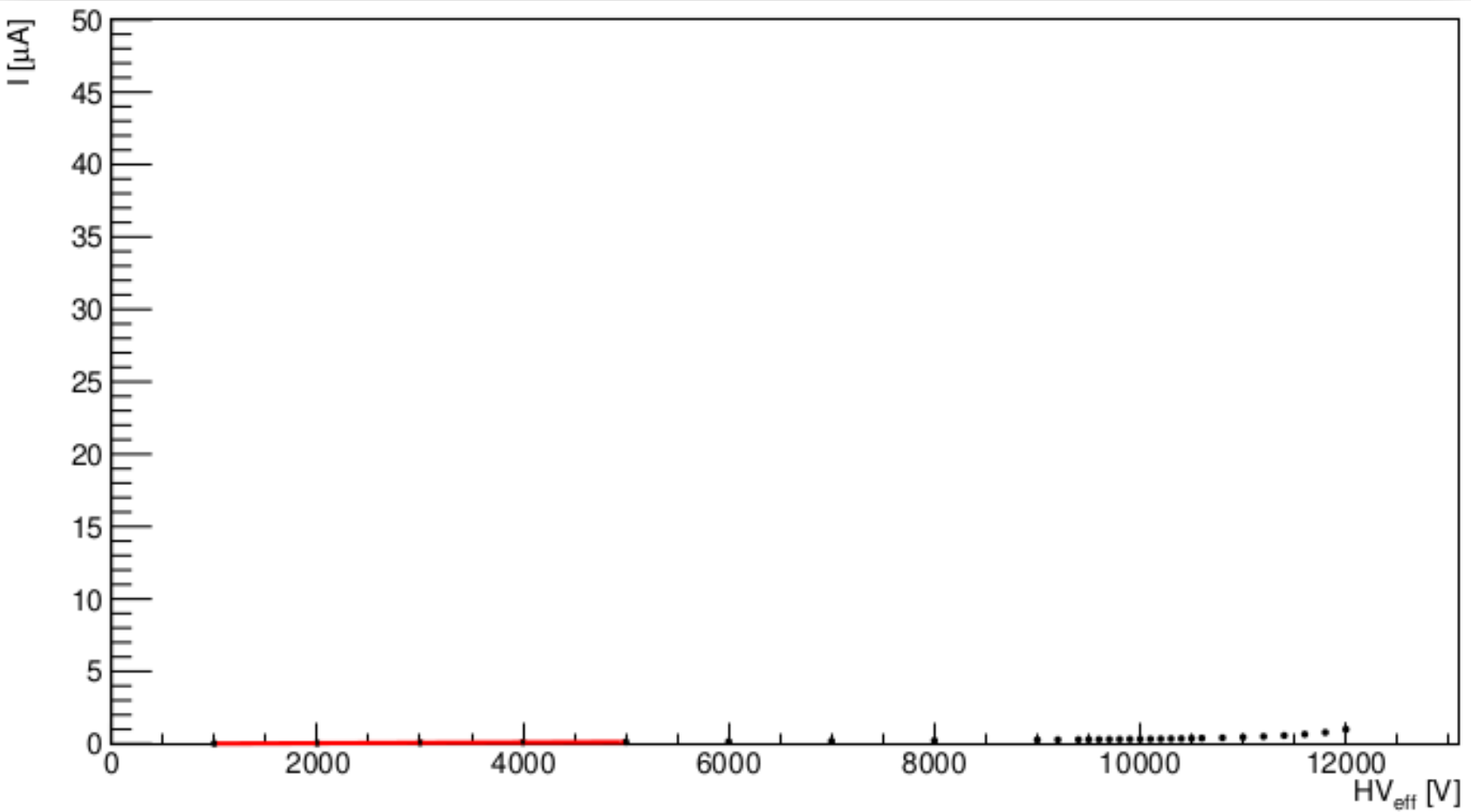}}
	\caption{Dark current drawn by the ALICE RPC as a function of the HV before the start of the irradiation, while flushed with ECO1 mixture. \label{ECO1_firstHVscan}}
\end{figure}

\begin{figure}[tb]
	\centerline{\includegraphics[width=8.5cm]{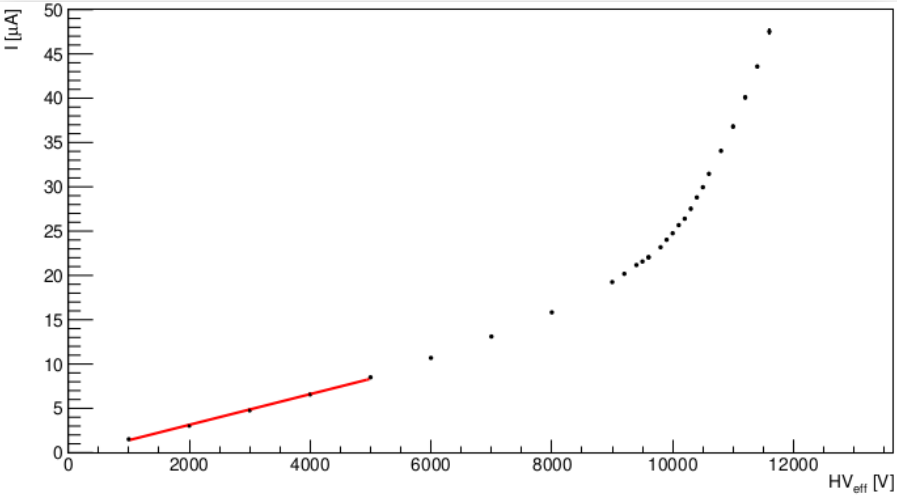}}
	\caption{Dark current drawn dy the ALICE RPC as a function of the HV after the integration of 22 mC/cm$^2$, while flushed with ECO1 mixture. \label{ECO1_lastHVscan}}
\end{figure}

The ageing test started in 2019, with two mixtures called ECO1 and ECO2. ECO1 is constituted by CO$_2$, HFO, {\it i}-C$_4$H$_{10}$, SF$_6$ in the proportion 50/45/4/1. ECO2 is constituted by the same gases in the proportion 60/35/4/1. The low percentage of isobutane allows to treat these mixture as non-flammable. \\ The chambers inside the GIF bunker are kept at a HV value close to the WP of the RPC and are irradiated for a prolonged amount of time (stability test). Every week a high voltage scan is performed without source in
order to measure the dark current (i.e. current with no source) and observe its long term behavior. \\ Dark current as a function of HV with the ECO1 gas mixture for the ALICE RPC are shown in the plots in figures \ref{ECO1_firstHVscan} and \ref{ECO1_lastHVscan}, respectively before and after the stability test (5 months between the two tests). We can observe an increase of the dark current. The fit used to estimate the Ohmic component of the dark current at 11.4 kV is shown in red. \\

\begin{figure}[tb]
	\centerline{\includegraphics[width=8.5cm]{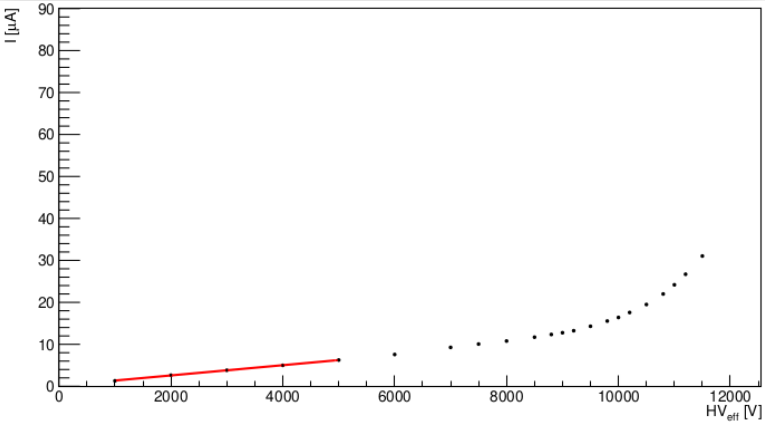}}
	\caption{Dark current drawn dy the ALICE RPC as a function of the HV before the start of the irradiation, while flushed with ECO2 mixture. \label{ECO2_firstHVscan}}
\end{figure}

\begin{figure}[tb]
	\centerline{\includegraphics[width=8.5cm]{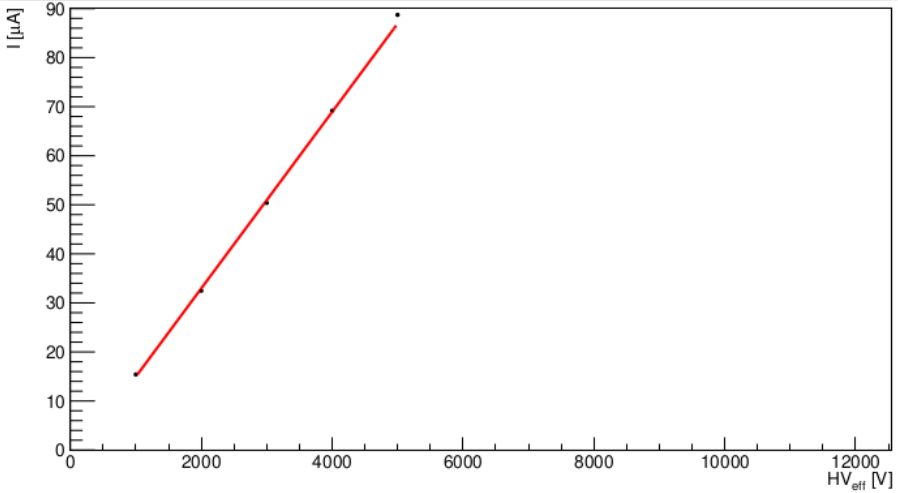}}
	\caption{Dark current drawn dy the ALICE RPC as a function of the HV after few days of short irradiation period, while flushed with ECO2 mixture. \label{ECO2_lastHVscan}}
\end{figure}

\newpage
In figures \ref{ECO2_firstHVscan} and \ref{ECO2_lastHVscan} the dark current vs. HV for the ECO2 gas mixture for the ALICE RPC, before and after the stability test, are shown. By increasing the percentage of CO$_2$, we have observed a large increase of the current as soon as we started the tests with ECO2 and the reasons for this increase are still under investigation (see Fig. \ref{ECO2_lastHVscan}). After that we had to remove the chamber since it was damaged: tests have continued with other RPCs from the collaboration and results for ECO2 are under investigation. 

\begin{figure}[tb]
	\centerline{\includegraphics[width=8.3cm]{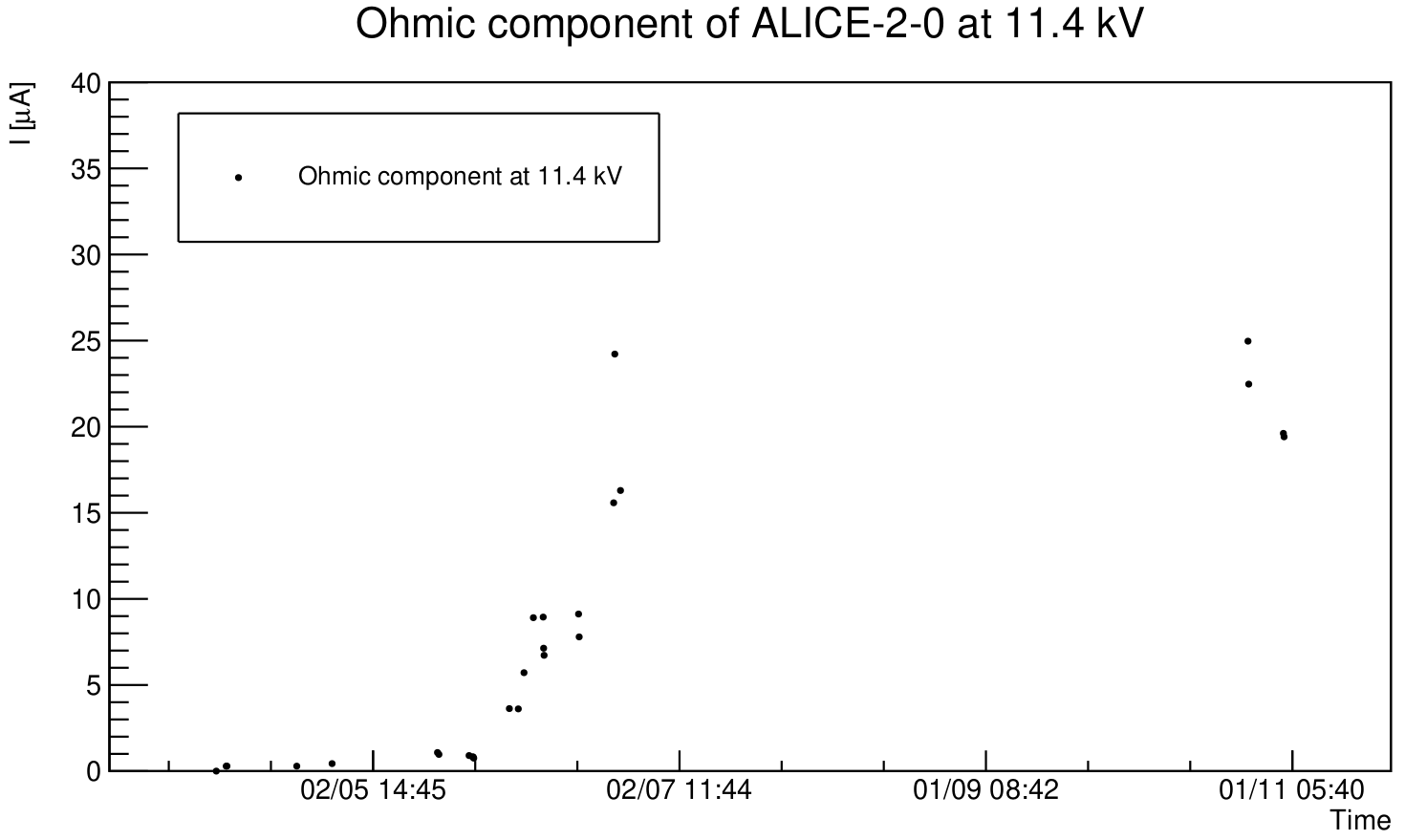}}
	\caption{Ohmic component of the current drawn by the ALICE RPC at the WP, while flushed with ECO1 mixture. \label{OhmicECO1}}
\end{figure}

\begin{figure}[tb]
	\centerline{\includegraphics[width=8.3cm]{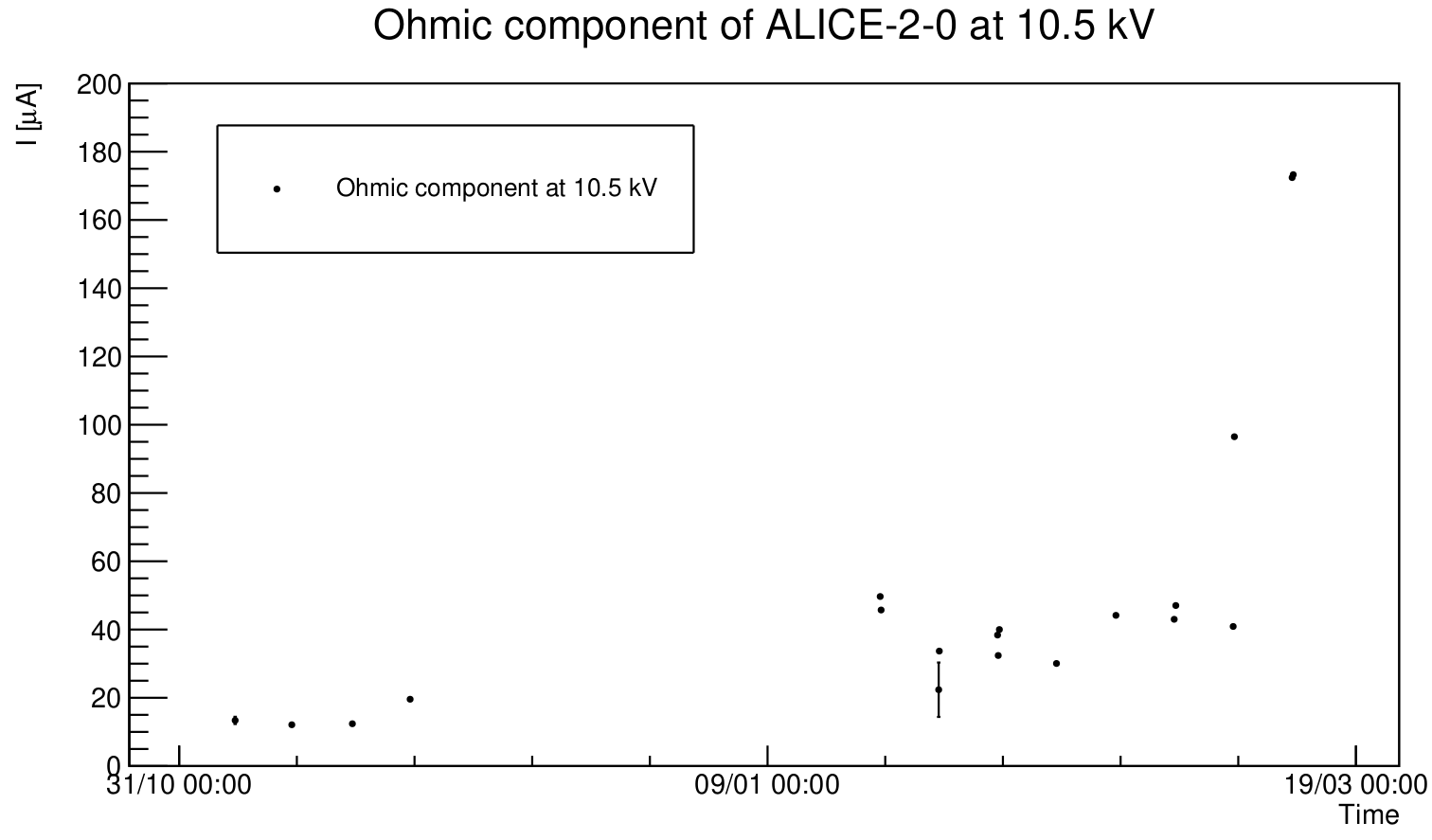}}
	\caption{Ohmic component of the current drawn by the ALICE RPC at the WP, while flushed with ECO2 mixture.  \label{OhmicECO2}}
\end{figure}

\newpage 
In the plots in figures \ref{OhmicECO1} and \ref{OhmicECO2} the Ohmic component (extrapolated from low HV linear fit from I-V curve) of the current for both the ECO1 and ECO2 gas mixture, at the RPC working point, is shown. Each point in the plots corresponds to an HV scan performed once a week, without source. The current increases over time for reasons still under investigation.

\newpage
\section{Conclusions}

Following R$\&$D studies on low GWP gas mixtures, C$_3$H$_2$F$_4$ seems a possible candidate to replace C$_2$H$_2$F$_4$. However the direct replacement of tetrafluoroethane with tetrafluoropropane is not possible because of the high WP ($>$ 14 kV). It has then been decided to add CO$_2$ to the gas mixture, in order to operate at lower voltages. Promising gas mixtures consisting of C$_3$H$_2$F$_4$, CO$_2$, {\it i}-C$_4$H$_{10}$ and SF$_6$, allow for a factor 5-20 GWP reduction with respect to the standard gas mixture, depending on the percentage variation of the different gases. \\
At GIF++ we have observed an increase of both ohmic and working current with the ECO gas mixtures under test: the link with CO$_2$ percentage in the gas mixture is under investigation. Tests beams have been performed in July, September and October 2021 at GIF++ with other RPC groups, in order to do more tests on ECO2 gas mixture, and these results are under analysis.

%\begin{thebibliography}{000} %for 3 digits
%\begin{thebibliography}{00}  %for 2 digits


\begin{thebibliography}{0}    %for 1 digit

%%journal paper
\bibitem{ALICE1} ALICE Collaboration, {\it The ALICE experiment at the CERN LHC}, Journal of Instrumentation (2008), 3.08: S08002

\bibitem{QGP1} S. Narison, {\it QCD as a theory of hadrons: from partons to confinement}, Cambridge, University Press (2004).

\bibitem{MS} {\it ALICE dimuon forward spectrometer: Addendum to the Technical Design Report.} Technical Design Report ALICE. CERN, Geneva (1999).

\bibitem{MID1} L. Terlizzi {\it The ALICE Muon IDentifier (MID)}, 2020 JINST {\bf 15} C10031

\bibitem{ALICERPC} R. Arnaldi {\it et al.}, {\it A low-resistivity RPC for the ALICE dimuon arm}, Nuclear Instruments and Methods in Physics Research Section A: Accelerators, Spectrometers, Detectors and Associated Equipment, Volume 451.2 (2000) 462-473.

\bibitem{Frontend} R. Arnaldi {\it et al.}, {\it Front-End Electronics for the RPCs of the ALICE Dimuon Trigger}, Nuclear Science, IEEE Transactions on, Volume 52 (2005).

\bibitem{Ageingtest1} R. Arnaldi {\it et al.}, {\it Ageing test of RPC for the Muon Trigger System for the ALICE experiment}, IEEE Symposium Conference Record Nuclear Science 2004., 2004, pp. 2072-2076 Vol. 4.

\bibitem{FEERIC1} S. Manen, P. Dupieux, B. Joly, F. Jouve and R. Vandaele, {\it FEERIC, a very-front-end ASIC for the ALICE muon trigger resistive plate chambers}, 2013 IEEE Nuclear Science Symposium and Medical Imaging Conference (2013 NSS/MIC), Seoul, 2013, pp. 1-4.

\bibitem{European} The European Parliament and the Council, {\it Regulation (EU) No 517/2014 on fluorinated greenhouse gases}, OJL 150 (2014), see pp. 195–230

\bibitem{IPCC} IPCC Climate, The physical science basis, {\it Contribution of working group I to the Fifth Assessment Report of the Intergovernmental Panel on Climate Change} 2014.

\bibitem{Guida} R. Guida, M. Capéans-Garrido and B. Mandelli, {\it Characterization of RPC operation with new environmental friendly mixtures for LHC application and beyond}, 2016 JINST 11 C07016.

\bibitem{Liberti} B. Liberti et al., {\it Further gas mixtures with low environment impact}, 2016 JINST 11 C09012.

\bibitem{Bianchi1} A. Bianchi, {\it R$\&$D studies on eco-friendly gas mixtures for the ALICE muon identifier}, 2019 JINST 14 C09003.

\bibitem{Bianchi2} A. Bianchi et al, {\it Studies on tetrafluoropropene-based gas mixtures with low environmental impact for Resistive Plate 	Chambers}, 2020 JINST 15 C04039.

\bibitem{FEERICALICE} M. Marchisone, {\it Performance of a resistive plate chamber equipped with a new prototype of amplified front-end electronics in the ALICE detector}, J. Phys. Conf. Ser. 889 (2017) 012011.

\bibitem{ageingALICE} ALICE collaboration, {\it Ageing tests on the low-resistivity RPC for the ALICE dimuon arm}, Nucl. Instrum. Meth. A 508 (2003) 106.

\bibitem{SF6} P. Camarri, R. Cardarelli, A. Di Ciaccio and R. Santonico, {\it Streamer suppression with SF-6 in RPCs operated in avalanche mode}, Nucl. Instrum. Meth. A 414 (1998) 317.

\bibitem{Quaglia} L. Quaglia et al., {\it Performance and aging studies for the ALICE muon RPCs}, JINST 16 (2020) C04002

\end{thebibliography}
\end{document}